\documentstyle[11pt,newpasp,psfig]{article}
\input ./journals.def

\newcommand{\msun}{\mbox{${\rm M}_\odot$}}
\newcommand{\rsun}{\mbox{${\rm R}_\odot$}}
\newcommand{\Rgc}{\mbox{${R_{\rm GC}}$}}
\def\apgt{\ {\raise-.5ex\hbox{$\buildrel>\over\sim$}}\ }
\def\aplt{\ {\raise-.5ex\hbox{$\buildrel<\over\sim$}}\ }

\begin{document}
\title{Life and Death of Young Dense Star Clusters near the Galactic Center}

\author{Simon F. Portegies Zwart}
\affil{Astronomical Institute `Anton Pannekoek', 
       University of Amsterdam, Kruislaan 403, the Netherlands and
       Institute for Computer Science, University of Amsterdam, Kruislaan 403 
}
\author{Stephen L.\ W.\ McMillan}
\affil{Department of Physics, Drexel University, Philadelphia, PA 19104, USA}

\author{Holger Baumgardt}
\affil{Institute of Advanced Physical and Chemical Research  RIKEN,
       2-1 Hirosawa, Wako-shi, Saitama 351-019, Japan}

\begin{abstract}
We discuss the structural change and degree of mass segregation of
young dense star clusters within about 100\,pc of the Galactic center.
In our calculations, which are performed with GRAPE-6, the equations
of motion of all stars and binaries are calculated accurately but the
external potential of the Galaxy is solved (semi)analytically.  The
simulations are preformed to model the Arches star cluster.  We find
that star clusters with are less strongly perturbed by the tidal field
and dynamical friction are much stronger affected by mass segregation;
resulting in a significant pile-up of massive stars in the cluster
center. At an age of about 3.5\,Myr more than 90 per cent of the stars
more massive than $\sim 10$\,\msun\, are concentrated within the
half-mass radius of the surviving cluster.  Star clusters which are
strongly perturbed by the tidal field of the parent Galaxy are much
less affected by mass segregation.
\end{abstract}

\section{Introduction}
In recent years a relatively new class of star clusters has been
discovered.  These systems are variously referred to in the literature
as young populous clusters, super star clusters, proto-globular
clusters, and even young globular clusters (although it remains
unclear if they are in any way related to the old globular clusters
observed in many galactic halos).  We prefer the term ``young dense
cluster'' (hereafter YDC, sometimes pronounced ``YoDeC'') because it
highlights the key defining properties of these compact stellar
systems.  The fact that these clusters are young means that stars of
all masses are still present, offering critical insights into the
stellar initial mass function and cluster structural properties at
formation.  The term ``dense'' means that dynamical evolution and
physical collisional processes can operate fast enough to compete with
and even overwhelm stellar evolutionary timescales.  Dense stellar
systems are places where wholly new stellar evolution channels can
occur, allowing the formation of stellar species completely
inaccessible by standard stellar and binary evolutionary pathways.

\begin{table*}[bp!]
\caption[]{\footnotesize Observed properties of selected young, dense
star clusters, mainly in or near the Milky Way.  Mass ($M$), age,
tidal radius ($r_{\rm tide}$) and half-mass radius ($r_{\rm hm}$) are
taken from the literature. }
\begin{flushleft}
\begin{tabular}{ll|rrrrrrr} \hline
Name   &ref&   $\log M/\msun$& $\log$(age/yr) & location 
		& $r_{\rm tide}$ (pc)& $r_{\rm hm}$ (pc) \\
\hline
Arches      &a& 4.8~~~~~ & 6.5~~~~~  & GC~   &   1~~  & 0.23~  \\
Quintuplet  &b& 4.2~~~~~ & 6.6~~~~~  & GC~   &   1~~  & 0.5~~   \\
\hline 
NGC\,3603   &c& 4.3~~~~~ & 6.5~~~~~  & disk~ &  10~~  & 0.78~  \\
Westerlund~1&d& 4.5~~~~~ & 6.8~~~~~  & disk~ &  10~~  & 0.2~~  \\
R\,136      &e& 4.7~~~~~ & 6.5~~~~~  & LMC~  &$>20$~~ & 0.5~~  \\
\hline
MGG-11      &f& 5.5~~~~~ & 7.0~~~~~  & M82~  &$>20$~~ & 1.2$^\star$~  \\ 
\hline
\end{tabular} \\
\medskip\footnotesize
References:
a) Figer et al.~(1999;2002;2004);
b) Glass et al.~(1987);
c) Brandl (1999);
d) Vrba et al.~(2000);
e) Brandl et al.~(1996);
f) McCrady et al. (2003)\\
$^\star$ The projected half light radius $r_{\rm hl}=1.2$\,pc, and
we take $r_{\rm hm} = \frac43 r_{\rm hl}$ (Spitzer
1987).
\end{flushleft}
\label{Tab:observed}
\end{table*}

Table \ref{Tab:observed} presents an overview of known YDCs in the
neighborhood of the Milky Way Galaxy, for which the relevant
parameters are well determined.  The clusters listed in the table are
selected on the basis of age ($\aplt 10$\,Myr) and half-mass
relaxation time ($\aplt 100$\,Myr); the latter criterion is not
universally used, but it best reflects the interplay between dynamics
and stellar evolution just described---clusters satisfying this
condition are expected to experience significant dynamical evolution
before their massive stars explode as supernovae.  All the clusters in
Table \ref{Tab:observed} thus lie in the regime where dynamical and
stellar evolution cannot be considered independently.

In the paper we concentrate on the Arches star cluster, which was
observed in detail by Figer et al (1999; 2002) and Stolte (2002).  The
$\sim 3$\,Myr old Arches cluster is located at a projected distance of
about 25\,pc from the Galactic center, has a half mass radius of about
0.23 parsec and contains between 20\,000 and $10^5$ stars. Together
with the Quintuplet star cluster (Glass 1987) they are the only young
dense star clusters which are strongly perturbed by the tidal field of
their parent Galaxy. McMillan et al (2004) discusses the internal
dynamical evolution and the possibility of the formation of an
intermediate mass black hole in the star cluster MGG-11 in the
starburst galaxy M82 (see also Portegies Zwart et al 2004), where
Baumgardt et al (2004) discusses the further consequences of the
presence of such a black hole.

\begin{figure}[htbp!]
\psfig{figure=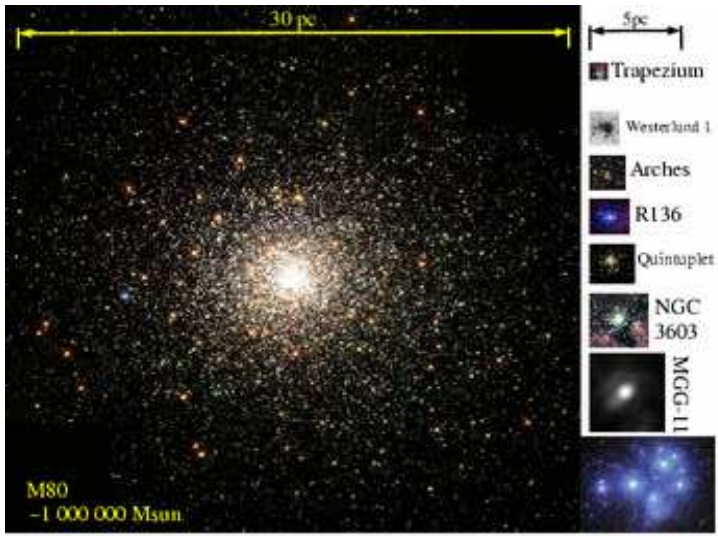,width=0.8\linewidth,angle=-90}
\caption[]{Composite image of several star clusters, all on the same
scale.  The big image to the left is the globular cluster M80. To the
right are a selection of young dense star clusters, identified
by their name (except the bottom image which represents the Pleiades
star cluster.  }
\label{Fig:composite}
\end{figure}

Figure~\ref{Fig:composite} shows a composite image of various star
clusters on the same scale. Among these are, next to some other YDCs
the Arches and Quintuplet systems. For comparison we added images of
the globular cluster M80, the Pleiades and the Trapezium cluster, the
star forming region in Orion.  We note that the latter three systems
are not considered YDCs, but are added for comparison.


We make the distinction between two families of YDCs; those that are
isolated and those that are strongly perturbed by the external tidal
field of their parent Galaxy.  We know only two clusters that are in
the latter category, Arches and Quintuplet, the other YDCs
are relatively unperturbed.  The evolution of tidally perturbed or
unperturbed clusters are profoundly different. In the limit of a
strong tidal field, stars near the tidal radius of the cluster feel
the proximity of the background Galaxy. 

\section{The simulation model and the initial conditions}

We simulate young dense star clusters in the tidal field of the
Galaxy.  As initial conditions we use 65536 stars from a Kroupa (2003)
initial mass function between the 0.1\,\msun\, and 100\,\msun, 16
percent of which receives a (secondary) companion star with a mass
between the adopted minimum stellar mass and the mass of the selected
(primary) star.  Orbital separations are selected from Roche-lobe
contact to 5kT (about 300\,\rsun), eccentricities are taken from the
thermal distribution.  The virial radius of all our models was
0.23\,pc, and the density profile was selected to be a King model with
$W_0=5$. These parameters are in agreement with the observed
parameters for the Arches cluster (Figer et al. 2002; Stolte et
al. 2002).  We adopt these current parameters as initial conditions
even though the cluster has experienced considerable dynamical
evolution over the last $\sim 3$\,Myr (Portegies Zwart et al 2002).

The star clusters are positioned in various orbits around the Galactic
center and evolved until an age of 3.5\,Myr. The age and orbital
parameters are selected such that the cluster then is at a distance of
about 30\,pc from the Galactic center.  We discuss the results of
three simulations here; (1) a star cluster (model R30a) in a circular
orbit at $\Rgc = 30$\,pc from the Galactic center, (2) a cluster which
is born in apo-Galacticon at 30\,pc from the Galactic center with 25
per cent of the velocity for a circular orbit (model R30b), and (2) a
cluster with a velocity of 15 percent of the circular velocity at an
apo-Galactic distance of 110\,pc (model R110).  In our calculations we
solve the equations of motion of all the stars in the cluster, the
orbit of the cluster and the evolution of the stars and binaries (see
Portegies Zwart, McMillan \& Gerhard, 2003, for details about the
choice of the tidal field). We use the GRAPE-6 (Makino et al. 1997;
2003) to speed up the calculations and the simulations ware run with
the {\tt Starlab} software environment (Portegies Zwart et al. 2001)
(see {\tt http://www.manybody.org/starlab/starlab.html}).

\section{Results}

\begin{figure}[htbp!]
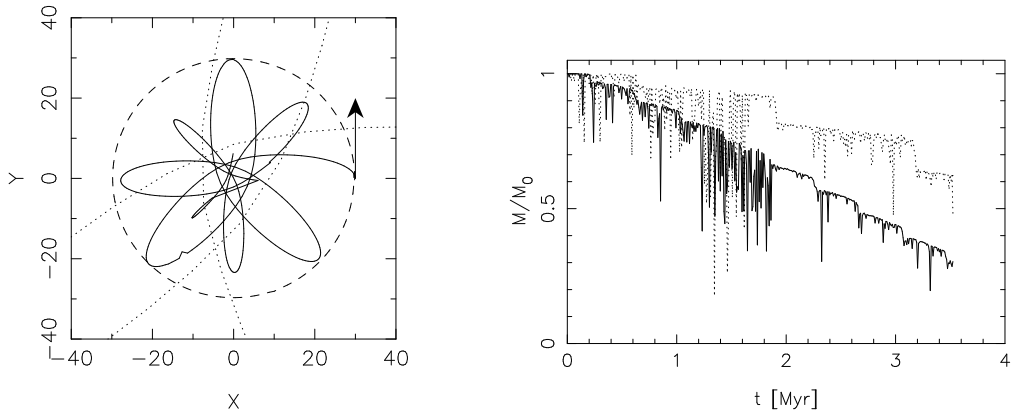

\begin{minipage}[b]{0.40\linewidth}
\psfig{figure=./Ar.N64.W5.MFK.r023.R30.orbit.ps,width=\linewidth,angle=-90}
\end{minipage}\hfill
\begin{minipage}[b]{0.50\linewidth}
\psfig{figure=./Ar.N64.W5.MFK.r023.Rvar.MvsT.ps,width=\linewidth,angle=-90}
\end{minipage}\hfill
\caption[]{ {\bf left:} Orbit of three simulated star clusters
($W_0=5$, $N=65536$, $r_{\rm vir} = 0.23$\,pc) one in a circular orbit
around the Galactic center (R30a: dashes), the other with a initial
orbital velocity of one quarter the circular velocity (R30b: solid
curve), the third cluster was born at a distance of 110\,pc from the
Galactic center with velocity 15\% of the circular velocity (R110:
dotted line). The trajectories are plotted until an age of 3.5\,Myr.
The arrow indicate the direction of the initial velocity of the
clusters born at $\Rgc = 30$\,pc.  \\
{\bf Right:} Evolution of the cluster mass for the models R30b (solid)
and R110 (dotted line) both are on elliptical orbits.  }
\label{Fig:orbit}
\label{Fig:massloss}
\end{figure}

Each star cluster, once initialized, evolves internally while it
orbits the Galactic center.  Figure\,\ref{Fig:orbit} (left frame)
shows the inner 40\,pc of the Galactic center with the orbits of three
star clusters until an age of 3.5\,Myr.  In the right frame of
figure\,\ref{Fig:massloss} we present the mass of the two clusters in
elliptic orbits as a function of time.  The cluster in the circular
orbit (R30a) in not plotted in this figure as it loses only $\sim 3$
per cent of its mass at a constant rate. The other two cluster on
elliptical orbits, lose mass at a much higher rate.  These clusters
lose mass predominantly near pericenter, while hardly any mass is lost
near apocenter. The average mass loss rates for these clusters is
about 7200\,\msun/Myr for model R30b and 4000\,\msun/Myr for model
R110, which are proportional to the relaxation time of these clusters
at the tidal radius at the moment of pericenter, consistent with the
expression derived by Portegies Zwart \& McMillan (2002).

The large (downward) spikes in the bound mass is mainly a result of
the fluctuating energy budget of the cluster due to binary evolution
and dynamical interactions involving binaries. Interesting to note is
that the binary fraction of the surviving clusters is about a factor
two higher than initially.

\begin{figure}[htbp!]
\begin{minipage}[b]{0.50\linewidth}
	\psfig{figure=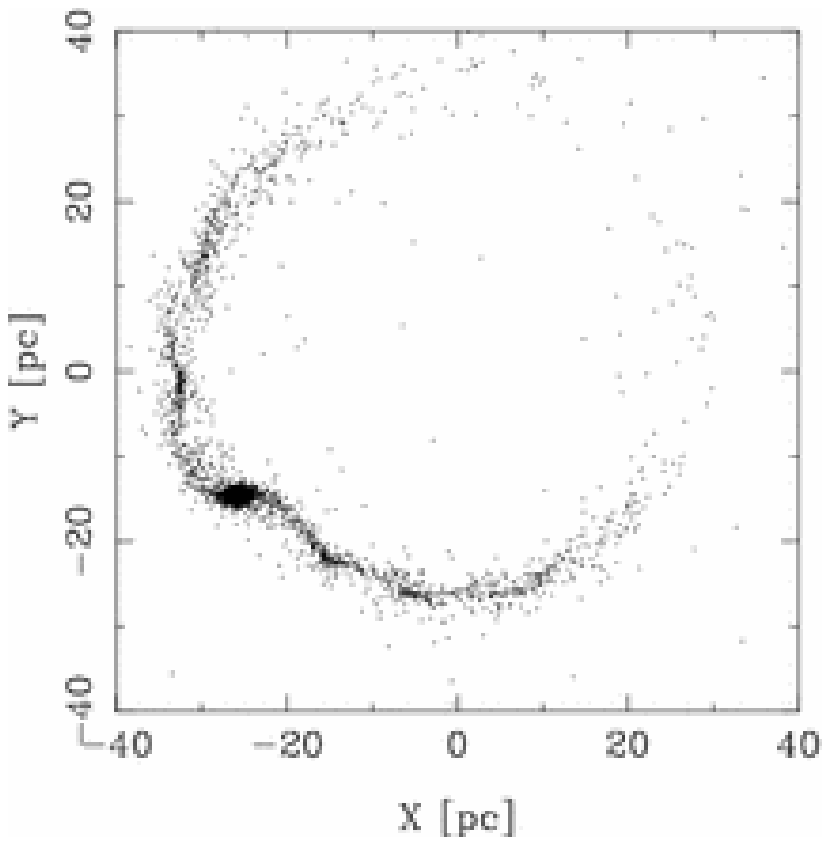,width=\linewidth,angle=-90}
        \psfig{figure=./Ar.N64.W5.MFK.r023.R30.v025.B16pc.T3.5Myr.XY.ps,width=\linewidth,angle=-90}
\end{minipage}\hfill
\begin{minipage}[b]{0.45\linewidth}
\psfig{figure=./Ar.N64.W5.MFK.r065.R110.T3.5Myr.XY.ps,width=\linewidth,angle=-90}
\end{minipage}\hfill
\caption[]{
Stellar positions at an age of about 3.5\,Myr for three simulated star
clusters ($W_0=5$, $N=65536$, $r_{\rm vir} = 0.23$\,pc).\\
{\bf Left-Top}: Calculation (R30a) with an initial circular orbit
starting a $\Rgc = 30$\,pc.  This cluster has orbited the Galactic
center more than three times in 3.5\,Myr. Most ($\sim 97$ per cent)
stars are still cluster members, and the escaped stars remain on more
or less the same orbit as the cluster.\\
{\bf Left-Bottom}: Calculation (R30b) in an initial orbit with 25\% of
the circular velocity and $\Rgc = 30$\,pc.  The small velocity has as
a consequence that the cluster passed the Galactic center at small
distance, resulting in strong tidal perturbations to the cluster. The
stars which became unbound at peri-Galacticon persist in their orbit,
whereas the orbit of the cluster is deflected by dynamical friction.\\
{\bf Right}: Calculation (R110) with $\Rgc = 110$\,pc and with a
velocity 0.15 of the circular velocity.  Last peri-Galacticon passage
occurred at an age of about 3.2\,Myr.
}
\label{Fig:disruption}
\end{figure}

Figure\,\ref{Fig:disruption} depicts the stellar positions of the
three clusters at an age of about 3.5\,Myr for models R30a (top left),
R30b (bottom left) and model R110 (right).  At a distance of 30\,pc
from the Galactic center dynamical friction is rather inefficient and
cluster R30a hardly sinks to the Galactic center on this short time
scale (see also McMillan \& Portegies Zwart 2003).  The structure of
the tidal debris in model R30a is a result of the low velocity of the
escaped stars, which follow epicyclic motions around the co-moving
fourth and fifth Lagrangian points of the combined potential of the
cluster and the background Galaxy. These density enhancements persist
with time, but may be hard to observe as they contain mainly low mass
stars.  However, we encourage observers to investigate the
surroundings of the Arches and Quintuplet clusters to find evidence
for these 'Trojan' stars or other signatures of tidal debris.

\begin{figure}[htbp!]
\psfig{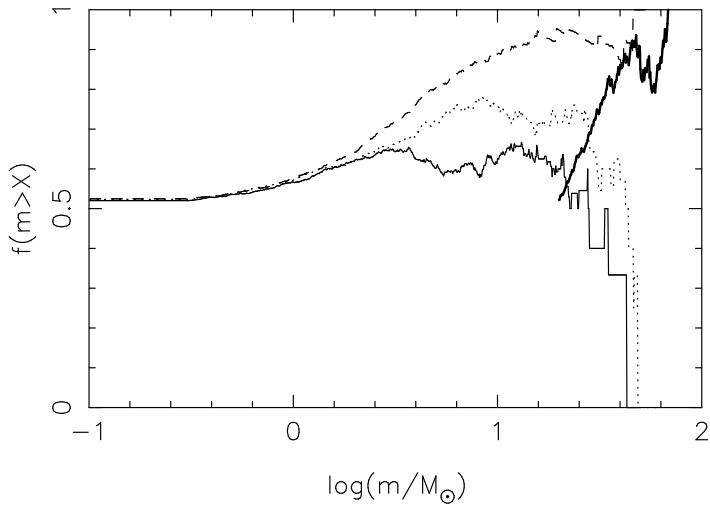}
\caption[]{Cumulative present day mass function within the surviving
cluster's half mass radius as fraction of the initial mass function.
The dashed curve is for the simulation in a circular orbit, the lower
solid curve is for the simulation at $\Rgc = 30$\,pc in an elliptic
orbit and the dotted line represents the star cluster with initial
$\Rgc = 110$\,pc in an elliptic orbit.  Thick solid line (upper right)
is taken from the Arches observations by Figer (2002) normalized to a
Scalo (1986) initial mass function.
}
\label{Fig:PDMF}
\end{figure}

The lower-left panel in Fig.\,\ref{Fig:disruption} shows the cluster
R30b, which was on an elliptic orbit. The obits of the stars which
escape from the cluster are slightly different than the orbit of the
cluster, which continues to be shocked again on each subsequent
passage, releasing even more stars.  The mass loss rate per pericenter
passage, however, is rather constant. Due to its rather rapid
evaporation this cluster is hardly affected by mass segregation (see
Fig.\,\ref{Fig:PDMF}).

The right panel in Fig.\,\ref{Fig:disruption} shows the cluster model
R110 which was born at a distance of 110\,pc from the Galactic center
with 15 per cent of the circular velocity, making the cluster orbit
quite radial.  The evolution of such cluster is relatively unperturbed
by the tidal field until first pericenter passage, which happens for
the first time around 0.6\,Myr.  The distribution of the stars at an
age of 3.5\,Myr is presented in Fig.\,\ref{Fig:disruption}.
Pericenter passage was reached some time earlier, at around 3.2\,Myr,
and the cluster now approaches apocenter again.

Figure\,\ref{Fig:PDMF} shows $f(m>X)$ the present day mass function
within the half-mass radius of the surviving cluster as fraction of
the initial mass function. The distributions are cumulative toward
higher mass.  For $m \sim 0.1$\,\msun\, this results in about $f(m>X)
\simeq 0.5$ confirming that we indeed compare the present day mass
function at about the half mass radius with the initial mass function.
This curve starts to deviate from $f(M>X) \simeq 0.5$ for higher mass
stars, indicating that these stars are overrepresented within the half
mass radius. The star cluster in a circular orbit at $\Rgc = 30$\,pc
(mode R30a) is most strongly affected by mass segregation. For
example, well over 90 per cent of the stars with $m \apgt 25$\,\msun\,
($\log(m/\msun) \apgt 1.4$) are present within the half-mass radius of
the surviving cluster.

The star cluster born at $\Rgc = 30$\,pc but with an elliptic orbit
(solid curve, model R30b) is least effected by mass segregation. This
is mainly caused by the rapid stripping of the cluster outer parts due
to the strong tidal field (see Fig.\,\ref{Fig:massloss}). The cluster
which was born at large distance form the Galactic center $\Rgc =
110$\,pc (model R110) is still strongly affected by mass segregation
even though the orbit brings it quite close to the Galactic center.
Around the time of pericenter passage the outer parts of the cluster
are stripped, but it takes about 0.6\,Myr to pass the Galactic center
for the first time.  This time is sufficient to allow a considerable
fraction of the high mass stars to segregate to the cluster center
which, at an age of 3.5\,Myr, is rich in high mass stars.

The thick solid curve in figure\,\ref{Fig:PDMF} represents the
observed excess of massive stars ($m>20$\,\msun) from the observations
of the Arches cluster by Figer et al.\, (2002), which was normalized
to the Scale (1986) mass function at the minimum mass of $\sim
20$\,\msun\, for which Figer (2002) claims to be complete. The
increasing incompleteness toward lower mass stars in these
observations makes the normalization to this extend hard, but we can
nicely compare the high-mass end which, for stars $\apgt 40$\,\msun\,
is rather consistent with cluster model R30a, and inconsistent with
the other simulations. From this comparison we conclude that the large
over-abundance of high-mass stars in the Aches cluster is consistent
with a mass function that is strongly affected by mass-segregation,
like in model R30a, but inconsistent with the weakly segregated mass
function of the strongly perturbed cluster R30e.

\section{Conclusions}
We perform simulations of young dense star clusters of 65536 stars of
which 10486 have a close binary companion. The clusters are initially
on three different orbits around the Galactic center. At an age of
3.5\,Myr we compare the mass function of the surviving cluster with
the observed present day mass function of the Arches cluster. From
this comparison we conclude that the present-day mass function of the
Arches cluster is highly dynamically evolved.  In the case this effect
is due to the dynamical evolution of the cluster, we conclude that the
cluster is not on an elliptic orbit which brings it within about
15\,pc from the Galactic center.

\section*{Acknowledgments}
We are grateful to Jun Makino and Piet Hut for numerous discussions,
and in particular to Jun for the excessive use of his GRAPE-6, on
which these simulations are performed. Additional simulations are
performed at the GRAPE-6 systems at Drexel University and at the
University of Amsterdam.  This work was supported by NASA ATP, the
Royal Netherlands Academy of Sciences (KNAW), the Dutch organization
of Science (NWO), and by the Netherlands Research School for Astronomy
(NOVA).

\bigskip
\bigskip
\noindent
{\bf References}
\bigskip


\noindent
Baumgardt, H, Portegies Zwart, S.F, McMillan, S.L.W, Makino, J, and Ebisuzaki, T, 2004, these proceedings

\noindent
Brandl, B., et al.\, {\em ApJ}, {\bf 352}, L69 (1999)

\noindent
Brandl, B., et al.\, {\em ApJ}, {\bf 466}, 254 (1996)

\noindent
{Figer}, D.~F, {McLean}, I.~S, {Morris}, M.~, {\em \apj} {\bf 514}, 202 (1999).

\noindent
{Figer}, D.~F. et~al., {\em \apj} {\bf 581}, 258 (2002).

\noindent
{Figer}, D.~F., 2004, these proceedings

\noindent
{Fuchs}, Y.~ et~al., {\em \aap} {\bf 350}, 891 (1999).

\noindent
{Glass}, I.~S, {Catchpole}, R.~M, {Whitelock},  P.~A. {\em \mnras} {\bf 227},
  373 (1987).

\noindent
{Kroupa}, P, {Weidner}, C.~{\em \apj} {\bf 598}, 1076 (2003).

\noindent
McCrady, N, Gilbert, A.M, Graham, J.R, {\em ApJ}, {\bf 596}, 240 (2003)

\noindent
{Makino}, J, {Taiji}, M, {Ebisuzaki}, T, {Sugimoto}, D.~{\em \apj} {\bf 480},
  432 (1997).

\noindent
{Makino}, J, {Fukushige}, T, {Koga}, M, {Namura}, K. {\em \pasj} {\bf 55}, 1163
  (2003).

\noindent
{McMillan}, S.~L.~W, {Portegies Zwart},  S.~F.~{\em \apj} {\bf 596}, 314 (2003).
\noindent
McMillan, S.L.W, Baumgardt, H, Portegies Zwart, S.F, 2004, these proceedings

\noindent
{Portegies Zwart}, S.~F, {Makino}, J, {McMillan}, S.~L.~W, {Hut}, P.~{\em \apj}
  {\bf 565}, 265 (2002).

\noindent
{Portegies Zwart}, S.~F, {McMillan},  S.~L.~W, {Gerhard}, O.~{\em \apj} {\bf
  593}, 352 (2003).

\noindent
{Portegies Zwart}, S.~F, {McMillan}, S.~L.~W, {Hut}, P, {Makino}, J.~{\em
  \mnras} {\bf 321}, 199 (2001).

\noindent
{Portegies Zwart} S.~F, Baumgardt, H., Hut, P., Makino, J., {McMillan}, S.~L.~W., {\em Nature} in press (astro-ph/0402622)

\noindent
Spitzer, L, in {\em Dynamical evolution of globular clusters}, 
Princeton University Press, p.\,191 (1987)

\noindent
{Stolte}, A.~, {Grebel}, E.~K., {Brandner}, W, {Figer}, D.~F. {\em \aap} {\bf
  394}, 459 (2002).

\noindent
{Vrba}, F.~J. et~al., {\em \apjl} {\bf 533}, L17 (2000).


\end{document}